\documentclass[aps,english,prl,amsmath,amsfonts,amssymb,superscriptaddress,twocolumn,showpacs,citeautoscript]{revtex4-1}
\usepackage[T1]{fontenc}
\usepackage[latin9]{inputenc}
\usepackage{amsmath}
\usepackage{amssymb}
\usepackage{wasysym}
\usepackage{esint}
\usepackage{graphicx}
\usepackage{times}
\usepackage{nicefrac}
\usepackage{appendix}
\usepackage{textcase}
\usepackage{subfigure}
\usepackage{empheq}
\usepackage{color}
\usepackage{leftidx}
\usepackage{makecell}
\usepackage{stackrel}
\makeatletter
\@ifundefined{textcolor}{}
{%
 \definecolor{BLACK}{gray}{0}
 \definecolor{WHITE}{gray}{1}
 \definecolor{RED}{rgb}{1,0,0}
 \definecolor{GREEN}{rgb}{0,1,0}
 \definecolor{BLUE}{rgb}{0,0,1}
 \definecolor{CYAN}{cmyk}{1,0,0,0}
 \definecolor{MAGENTA}{cmyk}{0,1,0,0}
 \definecolor{YELLOW}{cmyk}{0,0,1,0}
}

\renewcommand{\vec}[1]{\mathbf{#1}}

\renewcommand{\b}{\beta}

\newcommand{\add}[1]{\if\a\b{{\color{red} #1}}\else{#1}\fi}

\renewcommand{\eqref}[1]{(\ref{eq:#1})}

\newcommand{\Eqref}[1]{Equation~\ref{eq:#1}}
\newcommand{\figref}[1]{Fig.~\ref{fig:#1}}
\newcommand{\Figref}[1]{Figure~\ref{fig:#1}}

\makeatother
\usepackage{babel}

\begin{document}
\title{Non-additivity of van der Waals forces on liquid surfaces}

\author{Prashanth S. Venkataram}
\affiliation{Princeton University, Department of Electrical Engineering, Princeton, New Jersey 08544, USA}
\author{Jeremy D. Whitton}
\affiliation{Princeton University, Department of Physics, Princeton, New Jersey 08544, USA}
\author{Alejandro W. Rodriguez}
\affiliation{Princeton University, Department of Electrical Engineering, Princeton, New Jersey 08544, USA}

\date{\today}

\begin{abstract}
  We present an approach for modeling nanoscale wetting and dewetting
  of liquid surfaces that exploits recently developed, sophisticated
  techniques for computing van der Waals (vdW) or (more generally)
  Casimir forces in arbitrary geometries. We solve the variational
  formulation of the Young--Laplace equation to predict the equilibrium
  shapes of fluid--vacuum interfaces near solid gratings and show that
  the non-additivity of vdW interactions can have a significant impact
  on the shape and wetting properties of the liquid surface, leading to
  very different surface profiles and wetting transitions compared to
  predictions based on commonly employed additive approximations, such
  as Hamaker or Derjaguin approximations.
\end{abstract}

\maketitle


Wetting and dewetting phenomena are ubiquitous in soft matter systems
and have a profound impact on many disciplines, including
biology~\cite{Prakash2012}, microfluidics~\cite{Geoghegan2003}, and
microfabrication~\cite{Chakraborty2010}. One problem of great interest
concerns the suspension of fluid films on or near structured surfaces
where, depending on the interplay of competing short-range molecular
or capillary forces (e.g. surface tension), gravity, and long-range
dispersive interactions (i.e. van der Waals or more generally, Casimir
forces), the film may undergo wetting or dewetting transitions, or
exist in some intermediate state, forming a continuous surface profile
of finite thickness~\cite{Bonn2009, Geoghegan2003}. Thus far,
theoretical analyses of these competing effects have relied on
approximate descriptions of the dispersive van der Waals (vdW)
forces~\cite{arodreview, Israelachvili, Parsegian}, i.e. so-called
Derjaguin~\cite{Derjaguin1934} and Hamaker~\cite{Hamaker1937}
approximations, which have recently been shown to fail when applied in
regimes that fall outside of their narrow range of
validity~\cite{Buscher2004, lambrechtPWS, Emig2001,arodreview}.

\begin{figure}[t!]
\centering
\includegraphics[width=0.75\columnwidth]{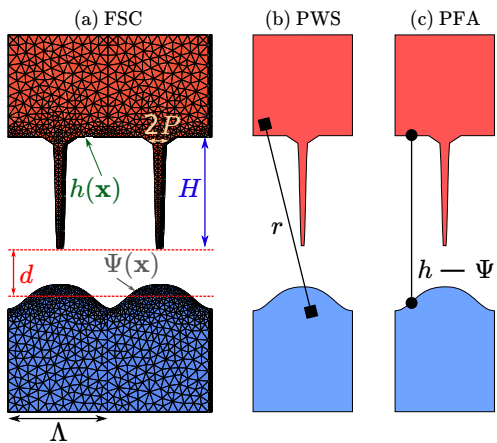}
\caption{Schematic of fluid--grating geometry comprising a fluid
  (blue) of surface profile $\Psi(\vec{x})$ in close proximity
  (average distance $d$) to a solid grating (red) of height profile
  $h(\vec{x})$, involving thin nanorods of height $H$, thickness $2P$,
  and period $\Lambda$. (a) Representative mesh employed by a recently
  developed FSC boundary-element method~\cite{SCUFF1} for computing
  exact vdW energies in complex geometries. (b) and (c) illustrate
  commonly employed pairwise--summation (PWS) and proximity--force
  approximations (PFA), involving volumetric and surface interactions
  throughout the bodies, respectively.}
\label{fig:schematic}
\end{figure}

In this paper, building on recently developed theoretical techniques
for computing Casimir forces in arbitrary geometries~\cite{Reid2013,
  Reid2009}, we demonstrate an approach for studying the equilibrium
shapes (the wetting and dewetting properties) of liquid surfaces that
captures the full non-additivity and non-locality of vdW
interactions~\cite{aroddesigner}. As a proof of concept, we consider
the problem of a fluid surface on or near a periodic grating,
idealized as a deformable perfect electrical conductor (PEC) surface
(playing the role of a fluid surface) interacting through vacuum below
a fixed periodic PEC grating [\figref{schematic}], and show that the
competition between surface tension and non-additive vdW pressure
leads to quantitatively and qualitatively different equilibrium fluid
shapes and wetting properties compared with predictions based on
commonly employed additive approximations. Our simplifying choice of
PEC surfaces allows for a scale-invariant analysis of the role of
geometry on both non-additivity and fluid deformations, ignoring
effects associated with material dispersion that would otherwise
further complicate our analysis and which are likely to result in even
larger deviations~\cite{Noguez2004,arodreview}. Our results provide a
basis for experimental studies of fluid suspensions in situations where
vdW non-additivity can have a significant impact.


Equilibrium fluid problems are typically studied by way of the
augmented Young-Laplace equation~\cite{interfacecolloidYLE},
\begin{equation}
  \gamma \nabla \cdot \left(\frac{\nabla \Psi}{\sqrt{1+|\nabla
        \Psi|^2}}\right) + \frac{\delta}{\delta\Psi}
  \left(\mathcal{E}_{\mathrm{other}}[\Psi]+\mathcal{E}_{\mathrm{vdW}}[\Psi]
  \right) =  0
\label{eq:YLE}
\end{equation}
describing the local balance of forces (variational derivatives of
energies) acting on a fluid of surface profile $\Psi(\vec{x})$. The
first two terms describe surface and other external forces (e.g.
gravity), with $\gamma$ denoting the fluid--vacuum surface tension,
while the third term $\frac{\delta}{\delta \Psi}
\mathcal{E}_{\mathrm{vdW}}$ denotes the local disjoining pressure
arising from the changing vdW fluid--substrate interaction energy
$\mathcal{E}_{\mathrm{vdW}}$. Semi-analytical~\cite{Quinn2013,
  Ledesma2012nanoscale} and brute-force~\cite{Ledesma2012multiscale,
  Sweeney1993} solutions of the YLE have been pursued in order to
examine various classes of wetting problems, including those arising
in atomic force microscopy, wherein a solid object (e.g. spherical
tip) is brought into close proximity to a fluid
surface~\cite{Quinn2013, Ledesma2012nanoscale, Ledesma2012multiscale},
or those involving liquids on chemically~\cite{Bauer1999, Checco2006}
or physically~\cite{Geoghegan2003, Bonn2009, Sweeney1993} textured
surfaces.

A commonality among prior theoretical studies of \eqref{YLE} is the
use of simple, albeit heuristic approximations that treat vdW
interactions as additive forces, often depending on the shape of the
fluid in a power-law fashion~\cite{Derjaguin1934, Derjaguin1956,
  Hamaker1937}. Derjaguin or proximity-force approximations (PFA) are
applicable in situations involving nearly planar structures,
i.e. small curvatures compared to their separation, approximating the
interaction between the objects as an additive, pointwise summation of
plate--plate interactions between differential elements comprising
their surfaces~\cite{Derjaguin1934, Derjaguin1956}.  Hamaker or
pairwise-summation (PWS) approximations are applicable in situations
involving dilute media~\cite{lambrechtPWS}, approximating the
interaction between two objects as arising from the pairwise summation
of (dipolar) London--vdW~\cite{caspol1} or
Casimir--Polder~\cite{caspol2} forces between volumetric elements of
the same constitutive materials~\cite{Hamaker1937}; such a treatment
necessarily neglects multiple-scattering and other non-additive
effects. When applied to geometries consisting of planar interfaces,
PFA can replicate exact results based on the so-called Lifshitz theory
(upon which it is based)~\cite{Dzyaloshinskii1961}, whereas PWS
captures the distance dependence obtained by exact calculations but
differs in magnitude (except in dilute situations)~\cite{lambrechtPWS}.
Typically, the quantitative discrepancy of PWS is rectified via a
renormalization of the force coefficient to that of the Lifshitz formula,
widely known as the Hamaker constant~\cite{Bergstrom1997}.



The inadequacy of these additive approximations in situations that
fall outside of their range of validity has been a topic of
significant interest, spurred by the recent development of techniques
that take full account of complicated non-additive and boundary
effects arising in non-planar structures, revealing non-monotonic,
logarithmic, and even repulsive interactions stemming from geometry
alone~\cite{arodreview, aroddesigner, arodpistons, bordagcyl}. These
brute-force techniques share little semblance with additive
approximations, which offer computational simplicity and intuition at
the expense of neglecting important electromagnetic effects. In
particular, the exact vdW energy in these modern formulations is often
cast as a log-determinant expression involving the full (no
approximations) electromagnetic scattering properties of the
individual objects, obtained semi-analytically or numerically by
exploiting spectral or localized basis expansions of the scattering
unknowns~\cite{arodreview, Lambrecht2006}. The generality of these
methods does, however, come at a price, with even the most
sophisticated of formulations requiring thousands or hundreds of
thousands of scattering calculations to be performed~\cite{arodreview}.
Despite the fact that fluid suspensions motivated much of the original
theoretical work on vdW interactions between macroscopic
bodies~\cite{Lamoreaux2006, Dzyaloshinskii1961, Israelachvili,
Parsegian}, to our knowledge these recent techniques have yet to be
applied to wetting problems in which non-additivity and boundary
effects are bound to play a significant role on fluid deformations.

\emph{Methods.--} In order to solve \eqref{YLE} in general settings,
we require knowledge of $\frac{\delta}{\delta\Psi}
\mathcal{E}_{\mathrm{vdW}} [\Psi]$ for arbitrary $\Psi$. We employ a
mature and freely available method for computing vdW interactions in
arbitrary geometries and materials~\cite{SCUFF1,SCUFF2}, based on the
fluctuating--surface current (FSC) framework~\cite{Reid2009, Reid2013}
of electromagnetic scattering, in which the vdW energy,
\begin{equation}
  \mathcal{E}_\mathrm{FSC} = \frac{\hbar}{2\pi}\int_0^\infty
  \mathrm{d}\xi \, \ln(\det(\mathbb{M} \mathbb{M}_{\infty}^{-1}))
\label{eq:FSC}
\end{equation}
is expressed in terms of ``scattering'' matrices $\mathbb{M}$,
$\mathbb{M}_\infty$ involving interactions of surface currents
(unknowns) flowing on the boundaries of the bodies~\cite{Reid2009,
  Reid2013} and integrated along imaginary frequencies $\xi =
\mathrm{i} \omega$; these are computed numerically via expansions in
terms of localized basis functions, or triangular meshes interpolated
by linear polynomials [\figref{schematic}(a)], in which case it is
known as a boundary element method. Because exact methods most commonly
yield the total vdW energy or force, rather than the local pressure on
$\Psi$, it is convenient to consider the YLE in terms of an equivalent
variational problem for the total energy~\cite{bormashenko,silin}:
\begin{equation}
  \min_{\Psi} \; \left(\gamma \int \sqrt{1 + |\nabla \Psi|^2} +
  \mathcal{E}_{\mathrm{other}}[\Psi] +
  \mathcal{E}_{\mathrm{vdW}}[\Psi]\right),
\label{eq:Emin}
\end{equation}
where just as in \eqref{YLE}, the first term captures the surface
energy, the second captures contributions from gravity or bulk
thermodynamic/fluid interactions, and the third captures the dispersive
vdW interaction energy. For simplicity, we ignore other competing
interactions, including thermodynamic and viscous
forces~\cite{Ledesma2012multiscale, Ledesma2012nanoscale} and neglect
gravity when considering nanoscale fluid deformations, focusing
instead only on the impact of surface and dispersive vdW interactions.

\Eqref{Emin} can be solved numerically via any number of available
nonlinear optimization/minimization
techniques~\cite{bormashenko,silin}, requiring only a convenient
parametrization of $\Psi$ using a finite number of degrees of
freedom. In what follows, we consider numerical solution of
\eqref{Emin} for the particular case of a deformable incompressible
PEC surface $\Psi$ interacting through vacuum with a 1d-periodic PEC
grating of period $\Lambda$ and shape $h(\vec{x}) = d -
H\left(\frac{1}{e^{\alpha (x - P)} + 1} + \frac{1}{e^{-\alpha (x + P)}
    + 1} - 2\right)$, for $|x| < \frac{\Lambda}{2}$, with half-pitch
$P = 0.03\Lambda$ and height $H =1.2\Lambda$. \Figref{schematic} shows
the grating surface and fluid profile obtained by solving \eqref{Emin}
for a representative set of parameters and mesh discretization. Here,
$d = 0.4\Lambda$ is the initial minimum grating-fluid separation, and
$\alpha\Lambda = 150$ is a parameter that smoothens otherwise sharp
corners in the grating, alleviating spatial discretization errors in
the calculation of $\mathcal{E}_{\mathrm{vdW}}$ while having a
negligible impact on the qualitative behavior of the energy compared
to what one might expect from more typical, piecewise-constant
gratings~\cite{Buscher2004}.

To minimize the energy, we employ a combination of algorithms found in
the NLOPT optimization suite~\cite{NLOPT, COBYLA, BOBYQA}.  Although
the localized basis functions or mesh of the FSC method provide one
possible parametrization of the surface, for the class of periodic
problems explored here, a simple Fourier expansion of the surface
provides a far more efficient and convenient basis, requiring far
fewer degrees of freedom to describe a wide range of periodic
shapes. Because the grating is translationally invariant along the $z$
direction and mirror-symmetric about $x = 0$, we parametrize $\Psi$ in
terms of a cosine basis, $\Psi(\vec{x}) = \sum_{n} c_n
\cos\left(\frac{2\pi{} nx}{\Lambda}\right)$, with the finite number of
coefficients $\{c_n\}$ functioning as minimization parameters. As we
show below, this choice not only offers a high degree of convergence,
requiring typically less than a dozen coefficients, but also
automatically satisfies the incompressibility or volume-conservation
condition $\int \Psi = 0$, which would otherwise require an
additional, nonlinear constraint. Note that the optimality and
efficiency of the minimization can be significantly improved when
local derivative information (with respect to the minimization
parameters) is available, but given that even a single evaluation of
$\mathcal{E}_{\mathrm{vdW}} [\Psi]$ is expensive---a tour-de-force
calculation involving hundreds of scattering
calculations~\cite{arodreview}---this is currently prohibitive in the
absence of an adjoint formulation (a topic of future
work)~\cite{Giles2000}. Given our interest in equilibrium fluid shapes
close to the initial condition of a flat fluid surface ($\Psi = 0$) and
because of the small number of degrees of freedom $\{c_n\}$ needed to
resolve the shapes, we find that local, derivative-free optimization is
sufficiently effective, yielding fast-converging solutions.

\begin{figure}[t!]
\centering
\includegraphics[width=0.97\columnwidth]{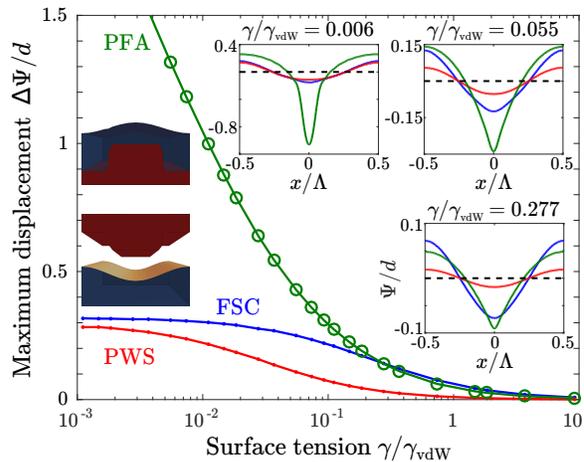}
\caption{Maximum displacement $\Delta\Psi/d$ of a fluid--vacuum
  interface that is repelled from a grating (insets) by a repulsive
  vdW force, as a function of surface tension
  $\gamma/\gamma_{\mathrm{vdW}}$, obtained via solution of
  \eqref{Emin} using FSC (blue), PWS (red), and PFA (green)
  methods. Circles indicate results obtained through \eqref{YLE}.
  Insets show the equilibrium fluid--surface profiles at selected
  $\gamma \in{} \{0.006, 0.055, 0.277\}\gamma_{\mathrm{vdW}}$, with
  the unperturbed $\Psi = 0$ surface denoted by black dashed lines.}
\label{fig:fig2}
\end{figure}

\begin{figure}[t!]
\centering
\includegraphics[width=0.87\columnwidth]{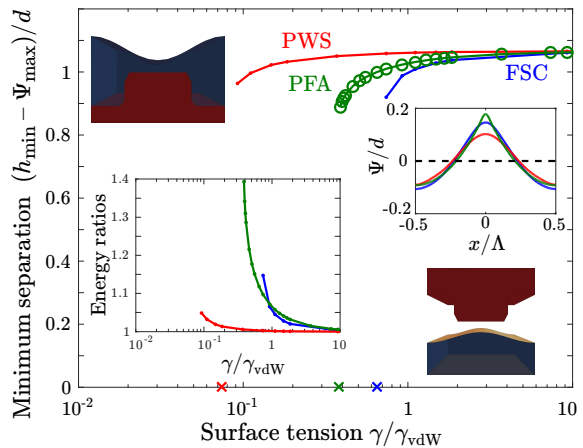}
\caption{Minimum surface--surface separation $\frac{h_{\mathrm{min}} -
    \Psi_{\mathrm{max}}}{d}$ of a fluid--vacuum interface that is
  attracted to a grating (insets) by an attractive vdW force, as a
  function of surface tension $\frac{\gamma}{\gamma_{\mathrm{vdW}}}$,
  obtained via solution of \eqref{Emin} using FSC (blue), PWS (red),
  and PFA (green) methods. Circles indicate results obtained through
  \eqref{YLE}.  Wetting transitions occurring at critical values of
  surface tension $\gamma^{(\mathrm{c})}$, marked as 'x'. The
  top-right inset shows the equilibrium fluid--surface profiles near
  $\gamma^{(\mathrm{c})}$ while the bottom-left inset shows the
  equilibrium vdW energies normalized by the energies of the
  unperturbed ($\Psi=0$) plane--grating geometry (the limit of $\gamma
  \to \infty$).}
\label{fig:fig3}
\end{figure}

In what follows, we compare the solutions of~\eqref{Emin} based
on~\eqref{FSC} against those obtained through PFA and PWS, which
approximate $\mathcal{E}_{\mathrm{vdW}}$ in this periodic geometry as:
\begin{align}
  \mathcal{E}_{\mathrm{PFA}} &= -\frac{\pi^2 \hbar c}{720}
  \int_{-\Lambda/2}^{\Lambda/2} \mathrm{d}x
  \left(\frac{1}{h(x) - \Psi(x)}\right)^{3} \label{eq:PFA} \\
  \mathcal{E}_\mathrm{PWS} &=
  A  \int_{-\Lambda/2}^{\Lambda/2} \mathrm{d}x'
  \int_{-\infty}^{\infty} \mathrm{d}x
  \int_{h(x')}^{\infty} \mathrm{d}y'
  \int_{-\infty}^{\Psi(x)} \mathrm{d}y \frac{1}{s^6},
\label{eq:PWS}
\end{align}
where $A = -\frac{2\pi\hbar c}{45}$ is a Hamaker-like coefficient
obtained by requiring that \eqref{PWS} yield the correct vdW energy
for two parallel PEC plates, as is typically
done~\cite{Bergstrom1997}. \Eqref{PWS} is obtained from pairwise
integration of the $r^{-7}$ Casimir--Polder interactions following
integration over $z$ and $z'$, with $r = \sqrt{s^2 + (z - z')^2}$ and
$s = \sqrt{(x - x')^2 + (y - y')^2}$~\footnote{Note that in situations
  involving a deformed PEC surface and flat PEC plate, one can show
  that $\mathcal{E}_\mathrm{PWS} =
  \mathcal{E}_\mathrm{PFA}$~\cite{Emig2003}, as this is a direct
  consequence of the additivity of the interaction.}. Note that
because we only consider perfect conductors, there is no dispersion to
set a characteristic length scale and hence all results can be quoted
in terms of an arbitrary length scale, which we choose to be
$\Lambda$. Additionally, we express the surface tension $\gamma$ in
units of $\gamma_{\mathrm{vdW}} = \frac{\pi^{2}\hbar c}{720d^{3}}$,
the vdW energy per unit area between two flat PEC plates separated by
distance $d$. In what follows, we consider the impact of
non-additivity on the fluid shape under both repulsive [\figref{fig2}]
or attractive [\figref{fig3}] vdW pressures (obtained by appropriate
choice of its sign), under the simplifying assumption of PEC surfaces
interacting through vacuum. In either case, we consider local
optimizations with small initial trust radii around $\Psi = 0$, and
characterize the equilibrium fluid profile $\Psi(x)$ as $\gamma$ is
varied. Our minimization approach is also validated against numerical
solution of \eqref{YLE} under PFA (green circles).


\emph{Repulsion.--} We first consider the effects of vdW repulsion on
the equilibrium profile of the fluid--vacuum interface, enforced in
our PEC model by flipping the sign of the otherwise attractive vdW
energy. Such a situation can arise when a fluid film either sits on or
is brought in close proximity to a solid grating
[\figref{fig2}(insets)], causing the fluid to either wet or dewet the
grating~\cite{Israelachvili}, respectively. \Figref{fig2} compares the
dependence of the maximum displacement $\Delta\Psi =
\Psi_{\mathrm{max}} - \Psi_{\mathrm{min}}$ of the fluid surface on
$\gamma$, as computed by FSC (blue), PWS (red), and PFA (green). Also
shown are selected surface profiles at small, intermediate, and large
$\gamma/\gamma_\mathrm{vdW}$. Note that the combination of a repulsive
vdW force, surface tension, and incompressibility leads to a
\emph{local} equilibrium shape that is corroborated via linear
stability analysis~\cite{ivanov1988thin}.


Under large $\gamma$, the surface energy dominates and thus all three
methods result in nearly-flat profiles, with $|\Psi| \ll d$. While
both additive approximations reproduce the exact energy of the
plane--plane geometry (with the unnormalized PWS energy
underestimating the exact energy by 20\%~\cite{lambrechtPWS}), we find
that (at least for this particular grating geometry)
$\mathcal{E}_\mathrm{PWS,PFA} / \mathcal{E}_\mathrm{FSC} \approx 0.25$
in the limit $\gamma \to \infty$, revealing that even for a flat fluid
surface, the grating structure contributes significant non-additivity.
Noticeably, at large but finite $\gamma \gg \gamma_\mathrm{vdW}$,
$\Delta\Psi$ is significantly larger under FSC and PFA than under PWS,
with $\Psi_\mathrm{FSC,PWS}$ exhibiting increasingly better qualitative
and quantitative agreement compared to the sharply peaked
$\Psi_\mathrm{PFA}$ as $\gamma$ decreases [\figref{fig2}(insets)]. The
stark deviation of PFA from FSC and PWS in the vdW--dominated regime
$\gamma \ll \gamma_{\mathrm{vdW}}$ is surprising in that PWS involves
volumetric interactions within the objects, whereas PFA and FSC depend
only on surface topologies. Essentially, the pointwise nature of PFA
means $\mathcal{E}_{\mathrm{PFA}}$ depends only on the local
surface--surface separation, decreasing monotonically with decreasing
separations and competing with surface tension and incompressibility
to yield a surface profile that nearly replicates the shape of the
grating in the limit $\gamma\to 0$. Quantitatively, PFA leads to
larger $\Delta\Psi$ as $\gamma \to 0$, asymptoting to a constant
$\lim_{\gamma \to 0} \Delta \Psi_{\mathrm{PFA}} \to H = 3d$ at
significantly lower $\frac{\gamma}{\gamma_{\mathrm{vdW}}} <
10^{-5}$. On the other hand, both $\mathcal{E}_{\mathrm{FSC}}$ and
$\mathcal{E}_{\mathrm{PWS}}$ exhibit much weaker dependences on the
fluid shape at low $\gamma$, with the former depending slightly more
strongly on the surface amplitude and hence leading to asymptotically
larger $\Delta\Psi$ as $\gamma \to 0$; in this geometry, we find that
$\Delta\Psi_{\mathrm{FSC,PWS}} \to \{0.32, 0.28\}d$ for
$\frac{\gamma}{\gamma_{\mathrm{vdW}}} \lesssim 10^{-2}$. Furthermore,
while PFA and PWS are found to agree with FSC at large and small
$\gamma$, respectively, neither approximation accurately predicts the
surface profile in the intermediate regime $\gamma \sim
\gamma_\mathrm{vdW}$, where neither vdW nor surface energies
dominate. Ultimately, neither of these approximations is capable of
predicting the fluid shape over the entire range of $\gamma$.


\emph{Attraction.--} We now consider the effects of vdW attraction,
which can cause a fluid film either sitting on or brought into
close proximity to a solid grating [\figref{fig3}(insets)] to dewet
or wet the grating, respectively~\cite{Israelachvili}. Here, matters
are complicated by the fact that $\mathcal{E}_\mathrm{vdW} \to -\infty$
as the fluid surface approaches the grating, leading to a fluid
instability or wetting transition below some critical
$\gamma^{(\mathrm{c})}$, depending on the competition between the
restoring surface tension and attractive vdW pressure. Such
instabilities have been studied in microfluidic systems through both
additive approximations~\cite{Bonn2009, kerle, Geoghegan2003,
Quinn2013}, but as we show in \figref{fig3}, non-additivity can lead
to dramatic quantitative discrepancies in the predictions obtained from
each method of computing $\mathcal{E}_{\mathrm{vdW}}$. To obtain
$\gamma^{(\mathrm{c})}$ along with the shape of the fluid surface for
$\gamma > \gamma^{(\mathrm{c})}$, we seek the nearest local solution
of~\eqref{Emin} starting from $\Psi = 0$. \Figref{fig3} quantifies the
onset of the wetting transition by showing the variation of the
minimum grating-fluid separation $h_{\mathrm{min}} -
\Psi_{\mathrm{max}}$ with respect to $\gamma$, as computed by FSC
(blue), PWS (red), and PFA (green), along with the corresponding
$\mathcal{E}_{\mathrm{vdW}}$ [\figref{fig3}(inset)] normalized to
their respective values for the plane--grating geometry (attained in
the limit $\gamma\to\infty$). Also shown in the top-right inset are
the optimal surface profiles at $\gamma\approx\gamma^{(\mathrm{c})}$
obtained from the three methods.

In contrast to the case of repulsion, here the fluid surface
approaches rather than moves away from the grating, which ends up
changing the scaling of $\mathcal{E}_{\mathrm{vdW}}$ with $\Psi$ and
leads to very different qualitative results. In particular, we find
that $\mathcal{E}_{\mathrm{FSC}}$ exhibits a much stronger dependence
on $\Psi_{\mathrm{max}}$ compared to PWS and PFA, leading to a much
larger $\gamma^{(\mathrm{c})}$ and a correspondingly broad surface
profile. As before, the strong dependence of
$\mathcal{E}_{\mathrm{PFA}}$ on the fluid surface, a consequence of
the pointwise nature of the approximation, produces a sharply peaked
surface profile, while the very weak dependence of
$\mathcal{E}_{\mathrm{PWS}}$ on the fluid shape ensures both a gross
underestimation of $\gamma^{(\mathrm{c})}$ along with a broader
surface profile. Interestingly, we find that
$\gamma^{(\mathrm{c})}_{\mathrm{FSC,PFA, PWS}} \approx
\{0.65,0.38,0.07\}\gamma_{\mathrm{vdW}}$, emphasizing the failure of
PWS to capture the critical surface tension by nearly an order of
magnitude.

\emph{Concluding Remarks.--} The predictions and approach described
above offer evidence of the need for exact vdW calculations for the
accurate determination of the wetting and dewetting behavior of fluids
on or near structured surfaces. While we chose to employ a simple
materials-agnostic and scale-invariant model for the vdW energy,
realistic (dispersive) materials can be readily analyzed within the
same formalism, requiring no modifications. We expect that in these
cases, non-additivity will play an even larger role. In fact, recent
works~\cite{Noguez2004, lambrechtPWS} have shown that additive
approximations applied to even simpler structures can contribute
larger discrepancies in dielectric as opposed to PEC bodies. For the
geometry considered above, assuming $\Lambda = 50 \, \mathrm{nm}$ and
a nonretarded Hamaker constant $A =
10^{-19}~\mathrm{J}$~\cite{Gu2001, Bergstrom1997, Israelachvili},
corresponding to a gold--water--oil material combination (with the thin
$d = 20~\mathrm{nm}$ water film taking the role of vacuum in our
model), we estimate that significant fluid displacements $\Delta \Psi
\sim 10~\mathrm{nm}$ and non-additivity can arise at $\gamma \approx
10^{-6}~\mathrm{J/m^{2}}$. By exploiting surfactants, it
should be possible to explore a wide range of $\gamma \in [10^{-7},
  10^{-2}] \, \mathrm{J/m^{2}}$~\cite{Quinn2013} and hence fluid
behaviors, from vdW- to surface-energy dominated regimes. Yet another
tantalizing possibility is that of observing these kinds of
non-additive interactions in extensions of the original liquid He$^4$
wetting experiments that motivated development of more general
theories of vdW forces (Lifshitz theory) in the first
place~\cite{Dzyaloshinskii1961}. In the future, it might also be
interesting to consider the impact of other forces, including but not
limited to gravity as well as finite-temperature thermodynamic effects
arising in the presence of gases in contact with fluid surfaces.

\emph{Acknowledgments.--} We are grateful to Howard A. Stone, M. T.
Homer Reid, and Steven G. Johnson for useful discussions. This
material is based upon work supported by the National Science
Foundation under Grant No. DMR-1454836 and by the National Science
Foundation Graduate Research Fellowship Program under Grant No. DGE
1148900.

\bibliography{wettingpaper}
\end{document}